\documentclass[a4paper, 12pt, openany, oneside]{article}
\usepackage[cp1251]{inputenc}
\usepackage[english]{babel}
\usepackage{ graphics,amsfonts, amsmath,bm,amssymb, array, eucal}
\usepackage[dvips]{graphicx}
\makeatletter
\renewcommand{\@biblabel}[1]{#1.\hfill}

\newcommand{\intl}{\int\limits}

\renewcommand{\Re}{\mathop{\rm Re\,}}
\renewcommand{\Im}{\mathop{\rm Im\,}}
\newcommand{\Res}{\mathop{\rm Res\,}}

\newcommand{\ch}{\mathop{\rm ch\, }}
\newcommand{\sh}{\mathop{\rm sh\, }}
\renewcommand{\th}{\mathop{\rm th\, }}
\makeatother {
\usepackage{indentfirst}

\begin{document}
\thispagestyle{empty} \large

\renewcommand{\refname}{\normalsize\begin{center}\bf
REFERENCES\end{center}}

\begin{center}
{\bf Landau's  Problem of Degenerate Plasma Oscillations in Slab
with Specular  Boundary Conditions}
\end{center}

\begin{center}
A. V. Latyshev\footnote{$avlatyshev@mail.ru$} and
  A. A. Yushkanov\footnote{$yushkanov@inbox.ru$}
\end{center}\medskip

\begin{center}
{\it Faculty of Physics and Mathematics,\\ Moscow State Regional
University,  105005,\\ Moscow, Radio st., 10--A}
\end{center}\medskip

\renewcommand{\abstractname}{\ }
\newcommand{\mc}[1]{\mathcal{#1}}
\newcommand{\E}{\mc{E}}

\thispagestyle{empty} \large

\date{\today}

\begin{abstract}
In the present paper the linearized problem of plasma oscillations
in slab (particularly, thin films) in external longitudinal 
alternating electric field is solved
analytically. Specular  boundary conditions of
electron reflection from the plasma boundary are considered.
Coefficients of continuous and discrete spectra of the problem are
found, and electron distribution function on the plasma boundary and
electric field are expressed in explicit form. Absorption of energy 
of electric field in slab is calculated.

Refs. 34. Figs. 2.\\

\end{abstract}


\begin{center}\bf
1. INTRODUCTION
\end{center}

The present paper is devoted to degenerate electron plasma behaviour
research. Analysis of processes taking place in plasma under effect
of external electric field, plasma waves oscillations with various
types of condi\-tions of electron reflection from the boundary has
important significance today in connection with problems of such
intensively developing fields as microelec\-tronics and nanotechnologies
\cite{1} -- \cite{Liboff}.

The concept of "plasma"\, appeared in the works of Tonks and Langmuir
for the first time (see \cite{Langmuir}--\cite{Tonks2}),
the concept of "plasma frequency"\, was introduced in the same works
and first questions of plasma oscillations were considered there.
However, in these works equation for the electric field was
considered separately from the kinetic equation.

A.A. Vlasov \cite{Vlasov} for the first time introduced the concept
of "self-consistent electric field"\, and added the corresponding item
to the kinetic equation. Now equations describing plasma behaviour
consist of anchor system of equations of Maxwell and Boltzmann.
The problem of electron plasma oscillations was considered by A.A.
Vlasov \cite{Vlasov} by means of solution of the kinetic equation
which included self-consistent electric field.

L.D. Landau \cite{Landau} had supposed that outside of the half-space
containing degenerate plasma external electromagnetic field causing
oscillations in plasma is situated. By this Landau has formulated a
boundary condition on the plasma boundary. After that the problem
of plasma oscillation turns out to be formulated correctly as a
boundary problem of mathematical physics.

In \cite{Landau}  L.D. Landau has solved analytically by Fourier
series the problem
of collisionless plasma behaviour in a half-space, situated in
external lon\-gi\-tu\-dinal (perpen\-di\-cular to the surface) electric
field, in conditions of specular reflection of electrons
from the boundary.

Further the problem of electron plasma oscillations was considered
by many authors. Full analytical solution of the problem is given
in the works \cite{8} and \cite{6}.

This problem has important significance in the theory of plasma (see,
for instance, \cite{2}, \cite{3} and the references in these works,
and also
\cite{M1}, \cite{M2}).

The problem of plasma oscillations with diffuse boundary condition
was considered in the works
\cite{Keller}, \cite{Kliewer} by method of integral transformations.
In the works \cite{Gohfeld1}, \cite{Gohfeld2} general asymptotic
analysis of electric field behaviour at the large distance from
the surface was carried out. In the work \cite{Gohfeld1} particular
significance of plasma behaviour analysis close to plasma re\-so\-nan\-ce
was shown. And in the same work \cite{Gohfeld1}  it was stated that
plasma behaviour in this case for conditions of specular and diffuse
electron scat\-te\-ring on the surface differs substantially.

In the works \cite{5} general questions of this problem
solvability were considered, but diffuse boundary conditions were
taken into account. In the work \cite{5} structure of discrete
spectrum in dependence of parameters of the problem was analyzed.
The detailed analysis of the solution in general case in the works
mentioned above hasn't been carried out conside\-ring the complex
character of this solution.

The present work is a continuation of electron plasma behaviour in
external longitudinal alternating electric field research
\cite{5} -- \cite{11}.

In the present paper the linearized problem of plasma oscillations
in slab in external alternating electric field is solved analytically.
Specular  boundary conditions for electron
reflection from the boundary are considered. In \cite{9}--\cite{11}
diffuse boundary conditions were considered.

The coefficients of continuous and discrete spectra of the problem
are obtained in the present work, which allows us to derive
expressions for electron distribution function at the boundary
of conductive medium and electric field in explicit form.

The present work is a continuation of our work \cite{Gritsienko},
in which questions of half-space plasma waves reflection 
from the plane boundary
bounding degenerate plasma were considered.

Let us note, that questions of plasma oscillations are also
considered in nonlinear statement (see, for instance, the work
\cite{Stenflo},
\cite{Stenflo2}).

\begin{center}\bf
2. PROBLEM STATEMENT
\end{center}

Let degenerate plasma occupy a slab  (particularly, thin films) $-a<x<a$.

We take system of equations describing plasma behaviour.
As a  kinetic equation we take Boltzmann --- Vlasov $\tau$--model 
kinetic equation:
$$
\dfrac{\partial f}{\partial t}+\mathbf{v}\dfrac{\partial f}{\partial
\mathbf{r}}+e\mathbf{E}\dfrac{\partial f}{\partial \mathbf{p}}=
\dfrac{f_{eq}(\mathbf{r},t)-f(\mathbf{r},\mathbf{v},t)}{\tau}.
\eqno{(1.1)}
$$

Here $f=f(\mathbf{r},\mathbf{v},t)$ is the electron distribution
function, $e$ is the electron charge, $\mathbf{p}=m\mathbf{v}$
is the momentum of an electron, $m$ is the electron mass, $\tau$
is the character time between two collisions,
$\mathbf{E}=\mathbf{E}(\mathbf{r},t)$ is the self-consistent
electric field inside plasma,
$f_{eq}=f_{eq}(\mathbf{r},t)$ is the local equilibrium Fermi --- Dirac distribution function,
$ f_{eq}=\Theta(\E_F(t,x)-\E), $ where $\Theta(x)$ is the function
of Heaviside,
$$
\Theta(x)=\left\{\begin{array}{c}
                   1, \quad x>0, \\
                   0,\quad x<0,
                 \end{array}\right.
$$
$\E_F(t,x)=\frac{1}{2}mv_F^2(t,x)$ is the disturbed kinetic energy
of Fermi, $\E=\frac{1}{2}mv^2$ is the kinetic energy of electron.

Let us take the Maxwell equation for electric field
$$
{\rm div}\,{\mathbf{E}(\mathbf{r},t)}= 4\pi\rho(\mathbf{r},t).
\eqno{(1.2)}
$$
Here $\rho(\mathbf{r},t)$ is the charge density,
$$
\rho(\mathbf{r},t)=e\int (f(\mathbf{r},
\mathbf{v},t)-f_{0}(\mathbf{v})) \,d\Omega_F, \eqno{(1.3)}
$$
where
$$
d\Omega_F=\dfrac{2d^3p}{(2\pi\hbar)^3}, \qquad
d^3p=dp_xdp_ydp_z.
$$

Here $f_{0}$  is the undisturbed Fermi --- Dirac electron 
distribution function,
$$
f_{0}(\E)=\Theta(\E_F-\E),
$$
$\hbar$ is the Planck's constant, $\nu$ is the effective frequency
of electron collisions, $\nu=1/\tau$, $\E_F=\frac{1}{2}mv_F^2$ is
the undisturbed kinetic energy of Fermi, $v_F$ is the electron
velocity at the Fermi surface, which is supposed to be spherical.

We assume that external electric field outside the plasma is
perpendi\-cu\-lar to the plasma boundary and changes according to
the following law: $E_{0}\exp(-i\omega t).$

Then one can consider that self-consistent
electric field
$\mathbf{E}(\mathbf{r},t)$ inside plasma has one
$x$--component and changes only lengthwise the axis $x$:
$$
\mathbf{E}=\{E_x(x,t),0,0\}.
$$

Under this configuration the electric field is perpendicular
to the boun\-da\-ry of plasma, which is situated in the plane $x=0$.

We will linearize the local equilibrium Fermi --- Dirac 
distribution $f_{eq}$
in regard to the undisturbed distribution $f_0(\E)$:
$$
f_{eq}=f_{0}(\E)+[\E_{F}(x,t)-\E]\delta(\E_{F}-\E),
$$
where $\delta(x)$ is the delta -- function of Dirac.

We also linearize the electron distribution function $f$ in terms of
absolute Fermi --- Dirac distribution
$f_{0}(\E)$:
$$
f=f_0(\E)+f_1(x,\mathbf{v},t).
\eqno{(1.4)}
$$

After the linearization of the equations (1.1)--(1.3) with the help
of (1.4) we obtain the following system of equations:
$$
\dfrac{\partial f_1}{\partial t}+v_{x}\dfrac{\partial f_1}{\partial
x}+\nu f_1(x, \mathbf{v}, t)
=\delta(\E_{F}-\E) \big[e E_{x}(x,t)v_{x}+\nu[\E_{F}(x,t)-\E_{F}]\big] ,
\eqno{(1.5)}
$$
$$
\dfrac{\partial E_{x}(x,t)}{\partial x}=\dfrac{8\pi
e}{(2\pi\hbar)^{3}}\int f_1(x,\mathbf{v'}, t)d^3p' 
\eqno {(1.6)}
$$

From the law of preservation of number of particles
$$
\int f_{eq}d\Omega_F=\int f d\Omega_F
$$
we find:
$$
[\E_{F}(x,t)-\E_{F}]\int\delta(\E_{F}-\E)d^3p=\int f_1 d^3p.
\eqno{(1.7)}
$$

From the equation (1.5) it is seen that we should search for
the function $f_1$ in the form proportional to the delta -- function:
$$
f_1=\E_F \delta(\E_F-\E) H(x,\mu,t), \qquad \mu=\dfrac{v_x}{v}.
\eqno{(1.8)}
$$

The system of equations (1.5) and (1.6) with the help of (1.7) and (1.8)
can be transformed to the following form:
$$
\dfrac{\partial H}{\partial t}+v_{F}\mu \dfrac{\partial H}{\partial
x}+ \nu H(x,\mu,t)=
$$

$$
=\dfrac{ev_{F}\mu}{\E_{F}}E_{x}(x,t)+
\dfrac{\nu}{2}\int_{-1}^{1}H(x,\mu',t)d\mu',
$$

$$
\dfrac{\partial E_{x}(x,t)}{\partial x}=\dfrac{16\pi^{2}e\E_{F}m^2
v_{F}}{(2\pi \hbar)^3}\int_{-1}^{1}H(x,\mu',t)d\mu'.
$$

Further we introduce dimensionless function
$$
e(x,t)=\dfrac{ev_{F}}{\nu \E_{F}}E_{x}(x,t)
$$
and pass to dimensionless coordinate $x_1=x/l$ , where $\;l=v_F\tau$
is the mean free path of electrons, and we introduce dimensionless
time $t_1=\nu t$. We obtain the following system of equations:

$$
\dfrac{\partial H}{\partial t_1}+\mu \dfrac{\partial H}{\partial
x_1}+ \nu H(x_1,\mu,t_1)
=\mu e(x_1,t_1)+\dfrac{1}{2}\int_{-1}^{1}H(x_1,\mu',t_1)d\mu',
\eqno{(1.9)}
$$

$$
\dfrac{\partial e(x_1,t_1)}{\partial
x_1}=\dfrac{3\omega_{p}^{2}}{2\nu^{2}}\int_{-1}^{1}H(x_1,\mu',t_1)d\mu'.
\eqno{(1.10)}
$$

Here $\omega_{p}$ is the electron (Langmuir) frequency 
of plasma oscillations,
$$
\omega_p^2=\dfrac{4\pi e^2N}{m},
$$
$N$ is the numerical density (concentration), $m$ is the electron mass.

We used the following well-known relation for degenerate plasma for
the derivation of the equations (1.9) and (1.10)
$$
\Big(\dfrac{v_F m}{\hbar}\Big)^3=3\pi^2 N.
$$

Let $k$ to be a dimensional wave number, and let us introduce
dimen\-si\-on\-less wave number
$k_1=k\dfrac{v_F}{\omega_p}$, then
$kx=\dfrac{k_1x_1}{\varepsilon}$, where
$\varepsilon=\dfrac{\nu}{\omega_p}$. We introduce the quantity
$\omega_1=\omega\tau=\dfrac{\omega}{\nu}$.

\begin{center}\bf
3.  BOUNDARY CONDITIONS STATEMENT
\end{center}

Let us outline the time variable of the functions $H(x_1,\mu,t_1)$
and $e(x_1,t_1)$, assuming
$$
H(x_1,\mu,t_1)=e^{-i\omega_1t_1}h(x_1,\mu),
\eqno{(2.1)}
$$
$$
e(x_1,t_1)=e^{-i\omega_1t_1}e(x_1).
\eqno{(2.2)}
$$

The system of equations (1.9) and (1.10) in this case will be
transformed to the following form:
$$
\mu\dfrac{\partial h}{\partial x_1}+(1-i\omega_1)h(x_1,\mu)= \mu
e(x_1)+\dfrac{1}{2}\int\limits_{-1}^{1}h(x_1,\mu')d\mu',
\eqno{(2.3)}
$$
$$
\dfrac{de(x_1)}{dx_1}=\dfrac{3\omega_p^2}{2\nu^2}
\int\limits_{-1}^{1}h(x_1,\mu')d\mu'.
\eqno{(2.4)}
$$

Further instead of $x_1,t_1$ we write $x,t$. We rewrite the system
of equations (2.3) and (2.4) in the form:
$$
\mu\dfrac{\partial h}{\partial x}+z_0h(x,\mu)= \mu
e(x)+\dfrac{1}{2}\int\limits_{-1}^{1}h(x,\mu')d\mu',
\eqno{(2.5)}
$$
$$
\dfrac{de(x)}{dx}=\dfrac{3}{2\varepsilon^2}
\int\limits_{-1}^{1}h(x,\mu')d\mu'. \eqno{(2.6)}
$$
Here
$$
z_0=1-i\omega_1=1-\frac{\omega}{\nu}=1-i\omega\tau=1-i\Omega,\qquad
\Omega=\omega\tau.
$$

We consider the external electric field outside the plasma  is
perpendi\-cu\-lar to the plasma boundary and changes according to
the following law: $E_{0}\exp(-i\omega t)$.
This means that for the field inside plasma on the plasma
boundary the following condition is satisfied:
$$
e(-a_1)=e_s,\qquad e(+a_1)=e_s,
\eqno{(2.7)}
$$
where $a_1=a/l$ is the dimensionless width (depth) of slab. Further
we will write $a$ instead $a_1$.

We consider specular  boundary conditions. For
the function $h(x,\mu)$ this conditions will be written in 
the following form:
$$
h(-a,\mu)=h(-a,-\mu), \qquad 0<\mu<1,
\eqno{(2.8)}
$$
and
$$
h(a,\mu)=h(a,-\mu), \qquad -1<\mu<0.
\eqno{(2.9)}
$$

Condition of symmetry of boundary conditions (2.7) -- (2.9)
and the equation (2.6) mean, that electric field $e(x)$ and function 
$h(x,\mu)$ in the slab possess properties of symmetry
$$
e(x)=e(-x),\qquad h(x,\mu)=h(x,-\mu).
\eqno{(2.10)}
$$

The non-flowing condition for the particle (electric current) flow
through the plasma boundary means that
$$
\int\limits_{-1}^{1}\mu\,h(-a,\mu)\,d\mu=
\int\limits_{-1}^{1}\mu\,h(a,\mu)\,d\mu=0. 
$$

In accordace to (2.10) this condition it is carried out automatically.

The problem statement is completed. Now the problem consists in
finding of such solution of the system of equations (2.5) and (2.6),
which satisfies the boundary conditions  (2.7) and (2.8). Further, with
the use of the solution of the problem, it is required to built the
profiles of the distribution function of the electrons moving to the
plasma surface, and profile of the electric field.

\begin{center}\bf
5. SEPARATION OF VARIABLES AND CHARACTERISTIC SYSTEM
\end{center}

Application of the general Fourier method of the separation of
variables in several steps results in the following 
substitution \cite{Case}:
$$
h_\eta(x,\mu)=\exp(-\dfrac{z_0x}{\eta})\Phi_1(\eta,\mu)+
\exp(\dfrac{z_0x}{\eta})\Phi_2(\eta,\mu),
\eqno{(3.1)}
$$
$$
e_\eta(x)=\Big[\exp(-\dfrac{z_0x}{\eta})+\exp(\dfrac{z_0x}{\eta})\Big]
E(\eta),
\eqno{(3.2)}
$$
where $\eta$ is the spectrum parameter or the parameter of
separation, which is complex in general.

We substitute the equalities (3.1) and (3.2) into 
the equations (2.5) and (2.6).
We obtain the following characteristic system of equations:
$$
z_0(\eta-\mu)\Phi_1(\eta,\mu)=\eta\mu E(\eta)+\dfrac{\eta}{2}
\int\limits_{-1}^{1}\Phi_1(\eta,\mu')d\mu',
\eqno{(3.3)}
$$
$$
z_0(\eta+\mu)\Phi_1(\eta,\mu)=\eta\mu E(\eta)+\dfrac{\eta}{2}
\int\limits_{-1}^{1}\Phi_2(\eta,\mu')d\mu',
\eqno{(3.4)}
$$
$$
-\dfrac{z_0}{\eta}E(\eta)=\dfrac{3}{\varepsilon^2}\cdot \dfrac{1}{2}
\int\limits_{-1}^{1}\Phi_1(\eta,\mu')d\mu',
\eqno{(3.5)}
$$
$$
\dfrac{z_0}{\eta}E(\eta)=\dfrac{3}{\varepsilon^2}\cdot
\dfrac{\eta}{2}\int\limits_{-1}^{1}\Phi_2(\eta,\mu')d\mu'.
\eqno{(3.6)}
$$

Let us introduce the designations:
$$
\eta_1^2=\dfrac{\varepsilon^2z_0}{3},\qquad 
\varepsilon=\dfrac{\nu}{\omega_p}.
$$

From equations (3.5) and (3.6) we obtain
$$
\int\limits_{-1}^{1}\Big[\Phi_1(\eta,\mu)d\mu+\Phi_2(\eta,\mu)\Big]d\mu=0.
\eqno{(3.7)}
$$
Let us introduce the designations
$$
n(\eta)=\int\limits_{-1}^{1}\Phi_1(\eta,\mu)d\mu.
\eqno{(3.8)}
$$
From equation (3.5) we obtain that
$$
E(\eta)=-\dfrac{3}{2\varepsilon^2z_0}\eta n(\eta),
\eqno{(3.9)}
$$
whence
$$
n(\eta)=-\dfrac{2\varepsilon^2z_0}{3}\cdot
\dfrac{E(\eta)}{\eta},
$$
or
$$
n(\eta)=-2\eta_1^2\dfrac{E(\eta)}{\eta}.
$$

By means of equalities (3.7) - (3.9) we will rewritten
the equations (3.3) and (3.4)
$$
(\eta-\mu)\Phi_1(\eta,\mu)=\dfrac{E(\eta)}{z_0}(\mu\eta-\eta_1^2),
\eqno{(3.10)}
$$
$$
(\eta+\mu)\Phi_2(\eta,\mu)=\dfrac{E(\eta)}{z_0}(\mu\eta+\eta_1^2).
\eqno{(3.11)}
$$

Solution of the system (3.10) and (3.11) depends essentially on the
condition whether
the spectrum parameter $\eta$ belongs to the interval $-1<\eta<1$.
In connection with this the interval $-1<\eta<1$ we will call as
continuous spectrum of the characteristic system.

Let the parameter $\eta\in (-1,1)$. Then from the equations 
(3.10) and (3.11) in
the class of general functions we will find eigenfunction
corresponding to the continuous spectrum
$$
\Phi_1(\eta,\mu)=\dfrac{E(\eta)}{z_0}P\dfrac{\mu\eta-\eta_1^2}{\eta-\mu}+
g_1(\eta)\delta(\eta-\mu),
\eqno{(3.12)}
$$
$$
\Phi_2(\eta,\mu)=\dfrac{E(\eta)}{z_0}P\dfrac{\mu\eta+\eta_1^2}{\eta+\mu}+
g_2(\eta)\delta(\eta-\mu).
\eqno{(3.13)}
$$

In these equations (3.12) and (3.13) $\delta(x)$ is 
the delta--function of Dirac,
the symbol $Px^{-1}$ means the principal value of the integral under
integrating of the expression $x^{-1}$.

Substituting now (3.12) and (3.13) in the equations (3.5) and (3.6),
We receive the equations
$$
\dfrac{E(\eta)}{z_0}\Bigg[1-\dfrac{\eta}{2c}\int\limits_{-1}^{1}
\dfrac{\mu'\eta-\eta_1^2}{\mu'-\eta}d\mu'\Bigg]=-\dfrac{\eta}{2c}g_1(\eta),
$$
$$
\dfrac{E(\eta)}{z_0}\Bigg[1+\dfrac{\eta}{2c}\int\limits_{-1}^{1}
\dfrac{\mu'\eta+\eta_1^2}{\mu'+\eta}d\mu'\Bigg]=\dfrac{\eta}{2c}g_1(\eta),
$$
from which we obtain
$$
g_1(\eta)=-2\eta_1^2\dfrac{\lambda(\eta)}{\eta}E(\eta),\qquad
g_2(\eta)=-g_1(\eta)=2\eta_1^2\dfrac{\lambda(\eta)}{\eta}E(\eta).
\eqno{(3.14)}
$$

Here dispersive function is entered
$$
\lambda(z)=1-\dfrac{z}{2c}\int\limits_{-1}^{1}\dfrac{\mu z-\eta_1^2}
{\mu-z}d\mu,
\eqno{(3.15)}
$$
where
$$
c=\dfrac{\varepsilon^2z_0^2}{3}=z_0\eta_1^2.
$$

Functions (3.12) and (3.13) are  called eigen functions of the continuous
spectrum, since the spectrum parameter $\eta$ fills out the
continuum $(-1,+1)$ compactly. The eigen solutions of the given
problem can be found from the equalities (3.1) and (3.2).

The dispersion
function $\lambda(z)$ we express in the terms of
the Case dispersion function \cite{Case}:
$$
\lambda(z)=1-\dfrac{1}{z_0}+\dfrac{1}{z_0}\Big(1-
\dfrac{z^2}{\eta_1^2}\Big)\lambda_c(z),
$$
where
$$
\lambda_c(z)=1+\dfrac{z}{2}\int\limits_{-1}^{1}\dfrac{d\tau}{\tau-z}=
\dfrac{1}{2}\int\limits_{-1}^{1}\dfrac{\tau\,d\tau}{\tau-z}
$$
is the Case dispersion function  \cite{Case}.

The boundary values of the dispersion function from above and below
the cut (interval $(-1,1)$) we define in the following way:
$$
\lambda^{\pm}(\mu)=\lim\limits_{\varepsilon\to 0, \varepsilon>0}
\lambda(\mu\pm i \varepsilon), \qquad \mu\in (-1,1).
$$

The boundary values of the dispersion function from above and below
the cut are calculated according to the Sokhotzky formulas
$$
\lambda^{\pm}(\mu)=\lambda(\mu)\pm \dfrac{i \pi\mu}
{2\eta_1^2z_0}(\eta_1^2-\mu^2),\quad -1<\mu<1,
$$
from where
$$
\lambda^+(\mu)-\lambda^-(\mu)=\dfrac{i \pi}{\eta_1^2z_0}
\,\mu(\eta_1^2-\mu^2),
$$
$$
\dfrac{\lambda^+(\mu)+\lambda^-(\mu)}{2}=\lambda(\mu),\quad-1<\mu<1,
$$
where
$$
\lambda(\mu)=1+\dfrac{\mu}{2\eta_1^2z_0} \int\limits_{-1}^{1}
\dfrac{\eta_1^2-\eta^2}{\eta-\mu}\,d\eta,
$$
and the integral in this equality is understood as singular in terms
of the principal value by Cauchy. Besides that, the function
$\lambda(\mu)$ can be represented in the following form:
$$
\lambda(\mu)=1-\dfrac{1}{z_0}+
\dfrac{1}{z_0}\Big(1-\dfrac{\mu^2}{\eta_1^2}\Big)\lambda_c(\mu),
$$
$$
 \lambda_c(\mu)=1+\dfrac{\mu}{2}\ln\dfrac{1-\mu}{1+\mu}.
$$

Substituting relations (3.14) in (3.12) and (3.13), we will present
last expressions in the following form
$$
\Phi_1(\eta,\mu)=\dfrac{E(\eta)}{z_0}
\Big[P\dfrac{\mu\eta-\eta_1^2}{\eta-\mu}-2c\dfrac{\lambda(\eta)}
{\eta}\delta(\eta-\mu)\Big],
$$
and
$$
\Phi_2(\eta,\mu)=\dfrac{E(\eta)}{z_0}
\Big[P\dfrac{\mu\eta+\eta_1^2}{\eta+\mu}+2c\dfrac{\lambda(\eta)}
{\eta}\delta(\eta+\mu)\Big],
$$
or
$$
\Phi_1(\eta,\mu)=\dfrac{E(\eta)}{z_0}F(\eta,\mu),\qquad
\Phi_2(\eta,\mu)=\dfrac{E(\eta)}{z_0}F(-\eta,\mu),
$$
where
$$
F(\eta,\mu)=P\dfrac{\mu\eta-\eta_1^2}{\eta-\mu}-2c\dfrac{\lambda(\eta)}
{\eta}\delta(\eta-\mu),
$$
and
$$
F(-\eta,\mu)=P\dfrac{\mu\eta+\eta_1^2}{\eta+\mu}+2c\dfrac{\lambda(\eta)}
{\eta}\delta(\eta+\mu).
$$

It will be necessary for us the following relation of symmetry
$$
F(-\eta,-\mu)=-F(\eta,\mu).
$$

Let us notice, that eigen function $F(\eta,\mu)$ satisfies
to following condition of normalization
$$
\int\limits_{-1}^{1}F(\eta,\mu)d\mu=-2cP\dfrac{1}{\eta},
$$
and that analogous
$$
\int\limits_{-1}^{1}F(-\eta,\mu)d\mu=2cP\dfrac{1}{\eta}.
$$

So, eigen function of a continuous spectrum is constructed and
it is defined by equality
$$
h_\eta(x,\mu)=\Big[\exp\Big(-\dfrac{xz_0}{\eta}\Big)F(\eta,\mu)+
\exp\Big(\dfrac{xz_0}{\eta}\Big)F(-\eta,\mu)\Big]\dfrac{E(\eta)}{z_0},
$$
or, in explicit form,
$$
h_\eta(x,\mu)=\Bigg\{\Big[\exp\Big(-\dfrac{xz_0}{\eta}\Big)
\dfrac{\mu\eta-\eta_1^2}{\eta-\mu}+
\exp\Big(\dfrac{xz_0}{\eta}\Big)
\dfrac{\mu\eta+\eta_1^2}{\eta+\mu}\Big]-
$$
$$
-2c\dfrac{\lambda(\eta)}{\eta}\Big[\exp\Big(-\dfrac{xz_0}{\eta}\Big)
\delta(\eta-\mu)-\exp\Big(\dfrac{xz_0}{\eta}\Big)\delta(\eta+\mu)\Big]
\Bigg\}
\dfrac{E(\eta)}{z_0}.
$$
Let us replace exponents by hyperbolic functions and
also we will transform both square brackets from the previous expression. 
As a result we receive, that
$$
h_\eta(x,\mu)=\dfrac{2E(\eta)}{z_0}
\Bigg\{\ch\dfrac{xz_0}{\eta}\Bigg[ P\Big(
\dfrac{\mu\eta-\eta_1^2}{\eta-\mu}+\dfrac{\mu\eta+\eta_1^2}
{\eta+\mu}\Big)+
$$
$$
+2c\dfrac{\lambda(\eta)}{\eta}(-\delta(\eta-\mu)+
\delta(\eta+\mu))\Bigg]+\sh\dfrac{xz_0}{\eta}\Bigg[P
\Big(-\dfrac{\mu\eta-\eta_1^2}{\eta-\mu}+\dfrac{\mu\eta+\eta_1}
{\eta+\mu}\Big)+
$$
$$
+2c\dfrac{\lambda(\eta)}{\eta}(\delta(\eta-\mu)+\delta(\eta+\mu))\Bigg]
\Bigg\}.
$$

We will designate further
$$
\varphi(\eta,\mu)=F(\eta,\mu)+F(-\eta,\mu)=
P\Big(\dfrac{\mu\eta-\eta_1^2}{\eta-\mu}+\dfrac{\mu\eta+\eta_1^2}
{\eta+\mu}\Big)=
$$
$$
=2P\dfrac{\mu(\eta^2-\eta_1^2)}{\eta^2-\mu^2},
$$
and
$$
\psi(\eta,\mu)=-F(\eta,\mu)+F(-\eta,\mu)=P
\Big(-\dfrac{\mu\eta-\eta_1^2}{\eta-\mu}+\dfrac{\mu\eta+\eta_1}
{\eta+\mu}\Big)=
$$
$$
=2P\dfrac{\eta(\mu^2-\eta_1^2)}{\eta^2-\mu^2}.
$$

Thus, eigen function of a continuous spectrum it is possible
to present in the form of a linear combination of a hyperbolic sine and
kosine
$$
h_\eta(x,\mu)=\dfrac{2E(\eta)}{z_0}\Bigg\{\ch\dfrac{xz_0}{\eta}\Big[
P\dfrac{\mu(\eta^2-\eta_1^2)}{\eta^2-\mu^2}-c\dfrac{\lambda(\eta)}{\eta}
\big(\delta(\eta-\mu)-\delta(\eta+\mu)\big)\Big]-
$$
$$
-\sh\dfrac{xz_0}{\eta}\Big[P\dfrac{\eta(\mu^2-\eta_1^2)}{\eta^2-\mu^2}-
c\dfrac{\lambda(\eta)}{\eta}
\big(\delta(\eta-\mu)+\delta(\eta+\mu)\big)\Big]\Bigg\}.
$$

Let us notice, that eigen functions of a continuous spectrum it is possible
to present and in such form
$$
h_\eta(x,\mu)=\dfrac{E(\eta)}{z_0}\Bigg[\ch\dfrac{xz_0}{\eta}
\Big(F(\eta,\mu)+F(-\eta,\mu)\Big)+
$$
$$
+\sh\dfrac{xz_0}{\eta}\Big(-F(\eta,\mu)+F(-\eta,\mu)\Big)\Bigg].
\eqno{(3.16)}
$$

\begin{center}\bf
6. EIGENFUNCTIONS OF THE DISCRETE SPECTRUM
\end{center}

According to the definition, the discrete spectrum of the
characteristic equation is a set of zeroes of the dispersion
equation
$$
\dfrac{\lambda(z)}{z}=0. 
\eqno{(4.1)}
$$

We start to search zeroes of the equation (4.1). Let us take Laurent
series of the dispersion function:
$$
\lambda(z)=\lambda_\infty+\dfrac{\lambda_2}{z^2}+
\dfrac{\lambda_4}{z^4}+\cdots,\qquad |z|>1.
\eqno{(4.2)}
$$

Here
$$
\lambda_\infty \equiv\lambda(\infty)=
1-\dfrac{1}{z_0}+\dfrac{1}{3z_0\eta_1^2},
$$
$$
\lambda_2=-\dfrac{1}{z_0}\Big(\dfrac{1}{3}-\dfrac{1}{5\eta_1^2}\Big),
$$
$$
\lambda_4=-\dfrac{1}{z_0}\Big(\dfrac{1}{5}-\dfrac{1}{7\eta_1^2}\Big).
$$

We express these parameters through the parameters $\Omega$ and
$\varepsilon$:
$$
\lambda_\infty \equiv\lambda(\infty)=
\dfrac{2(\Omega-1)+i\varepsilon+(\Omega-1)(\Omega-1+i\varepsilon)}
{(\Omega+i\varepsilon)^2},
$$
$$
\lambda_2=-\dfrac{9+5i\varepsilon(\Omega+i\varepsilon)}{15(\Omega+
i\varepsilon)^2},
$$
$$
\lambda_4=-\dfrac{15+7i\varepsilon(\Omega+i\varepsilon)}{35(\Omega+
i\varepsilon)^2}.
$$

It is easy seen that the dispersion function (3.15) in collisional plasma
(i.e. when $\varepsilon>0$) in the infinity has the value which doesn't
equal to zero:
$\lambda_\infty=\lambda(\infty)\ne 0$.

Hence, the dispersion equation has infinity as a zero $\eta_i=\infty$,
to which the discrete eigensolutions of the given system correspond:
$$
h_\infty(x,\mu)=\dfrac{\mu}{z_0},\qquad\;e_\infty(x)=1.
$$

This solution is naturally called as mode of Drude. It describes the
volume conductivity of metal, considered by Drude
(see, for example, \cite{16}).

Let us consider the question of the plasma mode existence in
details. We find finite complex zeroes of the dispersion function.
We use the principle of argument. We take the contour (see Fig. 1)
$$
\Gamma_\varepsilon^+=\Gamma_R\cup \gamma_\varepsilon,
$$
which is passed in the positive direction and which bounds the
biconnected domain $D_R$. This contour consists of the circumference
$$
\Gamma_R=\{z:\quad |z|=R,\quad\;R=\dfrac{1}{\varepsilon},\quad
\varepsilon>0\},
$$
and the
contour $\gamma_\varepsilon$, which includes the cut $[-1,+1]$, and
stands at the distance of $\varepsilon$ from it.

Let us note that the dispersion function has not any poles in the
domain $D_R$.
Then according to the principle of argument the number \cite{17} of
zeroes $N$ in the domain $D_R$ equals to:
$$
N=\dfrac{1}{2\pi
i}\oint\limits_{\Gamma_\varepsilon}d\,\ln\lambda(z).
$$

\begin{figure}
\includegraphics{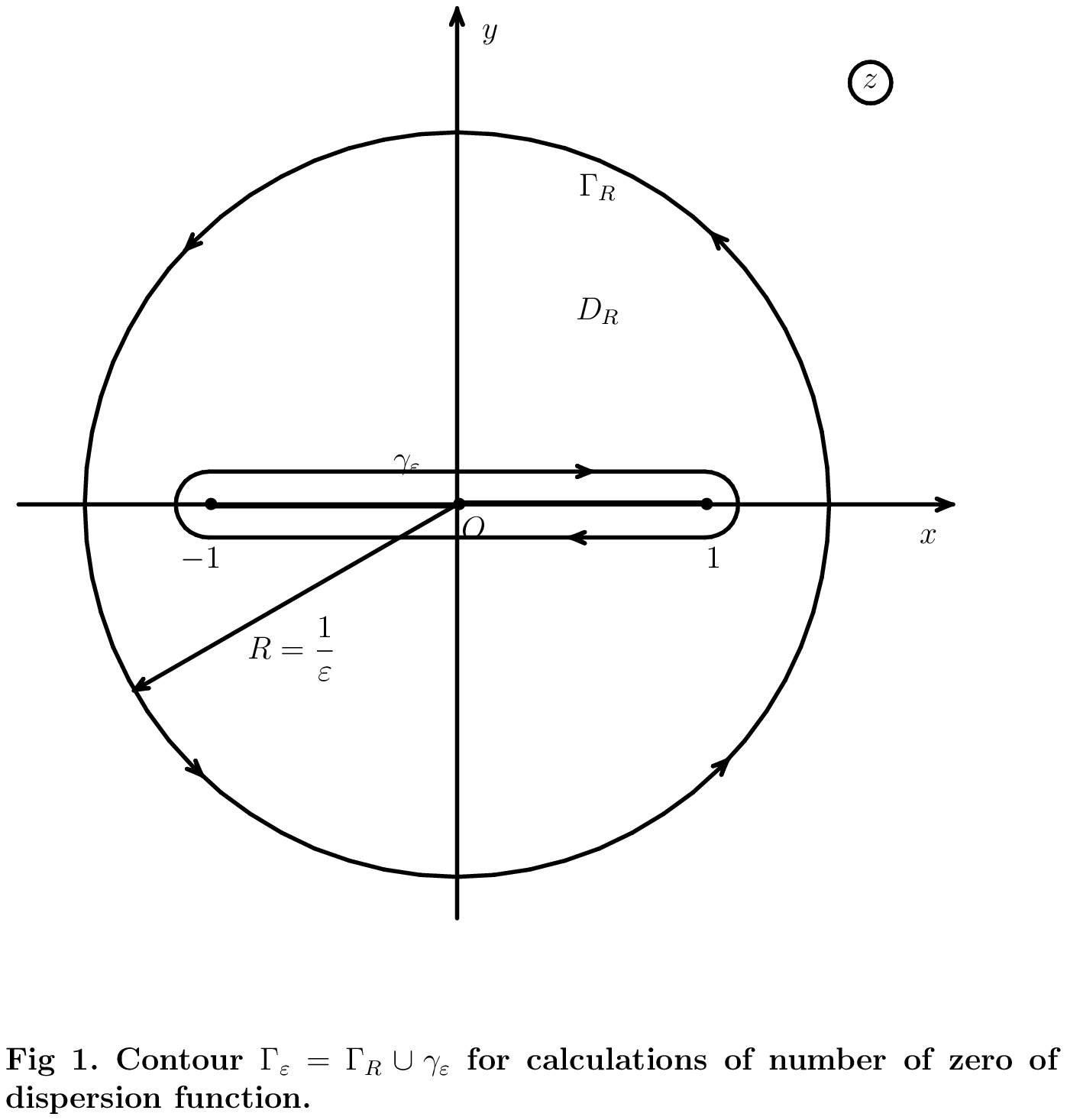}
\end{figure}

Considering the limit in this equality when $\varepsilon\to 0$ and
taking into account that the dispersion function is analytic in the
neighbourhood of the infinity, we obtain that
$$
N=\dfrac{1}{2\pi i}\int\limits_{-1}^{1}d\,\ln
\lambda^+(\tau)-\dfrac{1}{2\pi i}\int\limits_{-1}^{1}d\,\ln
\lambda^-(\tau)
$$
$$=\dfrac{1}{2\pi i}\int\limits_{-1}^{1}d\,\ln
\dfrac{\lambda^+(\mu)}{\lambda^-(\mu)}.
$$

So, we obtained that
$$
N=\dfrac{1}{2\pi i}\int\limits_{-1}^{1}d\,\ln
\dfrac{\lambda^+(\tau)}{\lambda^-(\tau)}.
$$
We divide this integral into two integrals by segments $[-1,0]$
and $[0,1]$. In the first integral by the segment $[-1,0]$ we
carry out replacement of variable
$ \tau \rightarrow -\tau $. Taking into account that
$\lambda^+(-\tau)=\lambda^-(\tau)$, we obtain that
$$
N=\dfrac{1}{2\pi i}\int\limits_{-1}^{1}d\,\ln
\dfrac{\lambda^+(\tau)}{\lambda^-(\tau)}= \dfrac{1}{\pi
i}\int\limits_{0}^{1}d\,\ln
\dfrac{\lambda^+(\tau)}{\lambda^-(\tau)}=\dfrac{1}{\pi}\arg
\dfrac{\lambda^+(\tau)}{\lambda^-(\tau)}\Bigg|_0^1.
\eqno{(4.3)}
$$

Here under symbol $\arg G(\tau)=\arg \dfrac{\lambda^+(\tau)}
{\lambda^-(\tau)}$ we understand the regular branch of the argument,
fixed in zero with the condition: $\arg G(0)=0$.

We consider the curve
$$
\gamma=\{z:\;z=G(\tau),\quad\;0\leqslant \tau\leqslant+1\},
$$
where
$$
G(\tau)=\dfrac{\lambda^+(\tau)}{\lambda^-(\tau)}.
$$

It is obvious that
$$
G(0)=1,\qquad\; \lim\limits_{\tau\to +1}G(\tau)=1.
$$

Consequently, according to (4.3), the number of values $N$ equals
to doubled number of turns of the curve $\gamma$ around the point
of origin, i.e.
$$
N=2\varkappa(G),
$$
where
$$
\varkappa(G)={\rm Ind}_{[0,+1]}G(\tau)
$$
is the index of the function $G(\tau)$.

Thus, the number of zeroes of the dispersion function, which are
situated in complex plane outside of the segment $[-1,1]$ of the
real axis, equals to doubled index of the function $G(\tau)$,
calculated on the "semi-segment"\, $[0,+1]$.

Let us single real and imaginary parts of the function $G(\mu)$ out.
At first, we represent the function $G(\mu)$ in the form:
$$
G(\mu)=\dfrac{(z_0-1)\eta_1^2+(\eta_1^2-\mu^2)\lambda_0(\mu)+
is(\mu)(\eta_1^2-\mu^2)}{(z_0-1)\eta_1^2+
(\eta_1^2-\mu^2)\lambda_0(\mu)- is(\mu)(\eta_1^2-\mu^2)}.
$$
where
$$
s(\mu)=\dfrac{\pi}{2}\mu, \qquad
$$
$$
\lambda(\mu)=1-\dfrac{1}{z_0}+
\dfrac{1}{z_0}\Big(1-\dfrac{\mu^2}{\eta_1^2}\Big)\lambda_c(\mu),
$$
and
$$
\lambda_c(\mu)=1+\dfrac{\mu}{2}\ln\dfrac{1-\mu}{1+\mu}
$$
is the dispersion function of Case, calculated on the cut
(i.e., in the interval $(-1,1)$).

Taking into account that
$$
z_0-1=-i\dfrac{\omega}{\nu}=-i\dfrac{\Omega}{\varepsilon}, \qquad
\eta_1^2=\dfrac{\varepsilon z_0}{3}=\dfrac{\varepsilon^2}{3}-
i\dfrac{\varepsilon \Omega}{3}, $$$$
(z_0-1)\eta_1^2=-\dfrac{\Omega^2}{3}-i\dfrac{\varepsilon \Omega}
{3},
$$
we obtain
$$
G(\mu)=\dfrac{P^-(\mu)+iQ^-(\mu)}{P^+(\mu)+iQ^+(\mu)},
$$
where
$$
P^{\pm}(\mu)=\Omega^2-\lambda_0(\mu)(\varepsilon^2-3\mu^2)\pm
\varepsilon \Omega s(\mu),
$$
$$
Q^{\pm}(\mu)=\varepsilon \Omega(1+\lambda_0(\mu))\pm
s(\mu)(\varepsilon^2-3\mu^2).
$$

Now we can easily single real and imaginary parts of the function
$G(\mu)$ out:
$$
G(\mu)=\dfrac{g_1(\mu)}{g(\mu)}+i\dfrac{g_2(\mu)}{g(\mu)}.
$$

Here
$$
g(\mu)=[P^+(\mu)]^2+[Q^+(\mu)]^2=[\Omega^2+\lambda_0(3\mu^2-
\varepsilon^2)-$$$$-\varepsilon \Omega s]^2+
[\varepsilon \Omega(1+\lambda_0)- s(3\mu^2-\varepsilon^2)]^2,
$$
$$
g_1(\mu)=P^+(\mu)P^-(\mu)+Q^+(\mu)Q^-(\mu)=[\Omega^2+
\lambda_0(3\mu^2-\varepsilon^2)]^2-$$$$\qquad\qquad-
\varepsilon^2 \Omega^2[s^2-(1+\lambda_0)^2]-
(3\mu^2-\varepsilon^2)^2s^2,
$$
$$
g_2(\mu)=P^+(\mu)Q^-(\mu)-P^-(\mu)Q^+(\mu)=
2s[\Omega^2(3\mu^2-\varepsilon^2)+$$$$
\qquad\qquad+\lambda_0(3\mu^2-
\varepsilon^2)^2+\varepsilon^2 \Omega^2(1+\lambda_0)],
$$

We consider (see Fig. 2) the curve $L$, which is defined in
implicit form by the following parametric equations:
$$
L=\{(\Omega,\varepsilon): \qquad g_1(\mu;\Omega,\varepsilon)=0,\;\quad
g_2(\mu;\Omega,\varepsilon)=0, \quad 0\le \mu\le 1\},
$$
and which lays in the plane of the parameters of the problem
$(\Omega,\varepsilon)$,
and when passing through this curve the index of the function $G(\mu)$
at the positive "semi-segment"\, $[0,1]$ changes stepwise.

From the equation $g_2=0$ we find:
$$
\Omega^2=-\dfrac{\lambda_0(\mu)(3\mu^2-\varepsilon^2)}
{3\mu^2+\varepsilon^2\lambda_0(\mu)}.
\eqno{(4.4)}
$$

Now from the equation $g_1=0$ with the help of (4.4) we find that
$$
\varepsilon=\sqrt{L_2(\mu)}, \eqno{(5.5)}
$$
where
$$
L_2(\mu)=-\dfrac{3\mu^2s^2(\mu)}{\lambda_0(\mu)
[s^2(\mu)+(1+\lambda_0(\mu))^2]}.
$$

Substituting (4.5) into (4.4), we obtain:
$$
\Omega=+\sqrt{L_1(\mu)}, 
\eqno{(4.6)}
$$
where
$$
L_1(\mu)=-\dfrac{3\mu^2[s^2(\mu)+
\lambda_0(\mu)(1+\lambda_0(\mu))]^2}{\lambda_0(\mu)[s^2(\mu)+
(1+\lambda_0(\mu))^2]}.
$$

Functions (4.5) and (4.6) determine the curve $L$ which is the
border if the domain $D^+$ (we designate the external area to the
domain as $D^-$) in explicit parametrical form (see Fig. 2). As in
the work \cite{18} we can prove that if $(\gamma,\varepsilon)\in
D^+$, then $\varkappa(G)={\rm Ind}_{[0,+1]} G(\mu)=1$ (the curve $L$
encircles the point of origin once), and if $(\gamma,\varepsilon)\in
D^-$, then $\varkappa(G)={\rm Ind}_{[0,+1]} G(\mu)=0$ (the curve $L$
doesn't encircle the point of origin).

We note, that in the work \cite{18}
the method of analysis of boundary regime when
$(\Omega,\varepsilon)\in L$ was developed.

From the expression (3.2) one can see that the number of zeroes of
the dispersion function equals to two if $(\Omega,\varepsilon)\in
D^+$, and equals to zero if $(\Omega,\varepsilon)\in D^-$.

\begin{figure}
\begin{center}
\includegraphics[width=16cm, height=10cm]{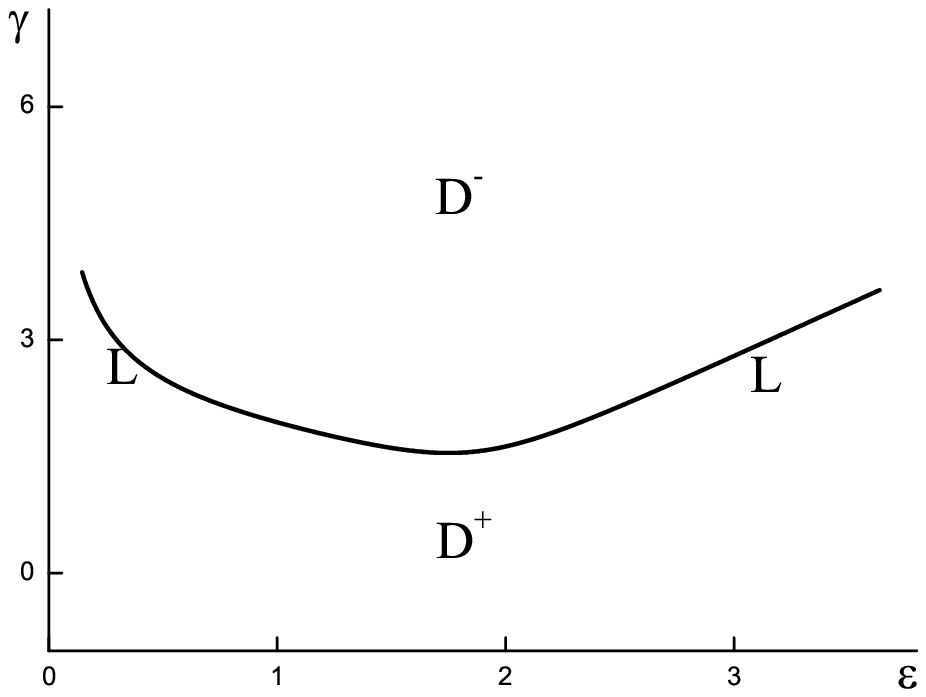}
\end{center}
\begin{center}
{\bf Fig. 2.}
\end{center}
\end{figure}

Since the dispersion function is even its zeroes differ from each
other by sign. We designate these zeroes as following $\pm \eta_0$,
by $\eta_0$ we take the zero which satisfies the condition $\Re
\eta_0>0$. The following solution corresponds to the zero $\eta_0$
$$
h_{\eta_0}(x,\mu)=\ch\dfrac{xz_0}{\eta_0}[F(\eta_0,\mu)+F(-\eta_0,\mu)]
+$$$$\hspace{3cm}
+\sh\dfrac{xz_0}{\eta_0}[-F(\eta_0,\mu)+F(-\eta_0,\mu)],
\eqno{(4.7)}
$$
$$
e_{\eta_0}(x)=2\ch\dfrac{z_0x}{\eta_0}.
\eqno{(4.8)}
$$

Here
$$
F(\eta_0,\mu)=\dfrac{\eta_0\mu-\eta_1^2}{\eta_0-\mu},\qquad
F(-\eta_0,\mu)=\dfrac{\eta_0\mu+\eta_1^2}{\eta_0+\mu}.
$$

It is easy to see, that function $h_{\eta_0}(x,\mu) $ is even
on $\eta_0$
$$
h_{\eta_0}(x,\mu)=h_{-\eta_0}(x,\mu).
$$

Function $h_{\eta_0}(x,\mu) $ we will present in the explicit form
$$
h_{\eta_0}(x,\mu)=\ch\dfrac{xz_0}{\eta_0}\Bigg[
\dfrac{\eta_0\mu-\eta_1^2}{\eta_0-\mu}+
\dfrac{\eta_0\mu+\eta_1^2}{\eta_0+\mu}\Bigg]+
$$
$$\hspace{3cm}
+\sh\dfrac{xz_0}{\eta_0}\Bigg[-
\dfrac{\eta_0\mu-\eta_1^2}{\eta_0-\mu}+
\dfrac{\eta_0\mu+\eta_1^2}{\eta_0+\mu}\Bigg],
\eqno{(4.9)}
$$
or
$$
h_{\eta_0}(x,\mu)=\ch\dfrac{xz_0}{\eta_0}\varphi(\eta_0,\mu)
+\sh\dfrac{xz_0}{\eta_0}\psi(\eta_0,\mu).
\eqno{(4.9')}
$$

Here
$$
\varphi(\eta_0,\mu)=F(\eta_0,\mu)+F(-\eta_0,\mu)=\hspace{6cm}
$$
$$\hspace{3cm}
=\dfrac{\eta_0\mu-\eta_1^2}{\eta_0-\mu}+
\dfrac{\eta_0\mu+\eta_1^2}{\eta_0+\mu}=\dfrac{2\mu(\eta_0^2-\eta_1^2)}
{\eta_0^2-\mu^2},
$$
$$
\psi(\eta_0,\mu)=-F(\eta_0,\mu)+F(-\eta_0,\mu)=\hspace{6cm}$$$$
$$
$$\hspace{3cm}
=-\dfrac{\eta_0\mu-\eta_1^2}{\eta_0-\mu}+
\dfrac{\eta_0\mu+\eta_1^2}{\eta_0+\mu}=\dfrac{2\eta_0(\eta_1^2-\mu^2)}
{\eta_0^2-\mu^2}.
$$

This solution is naturally called as mode of Debay (this is plasma
mode). In the case of low frequencies it describes well-known
screening of Debay \cite{Abrikosov}. The external field penetrates
into plasma on the depth of $r_D,\; r_D$ is the raduis of Debay.
When the external field frequencies are close to Langmuir
frequencies, the mode of Debay describes plasma oscillations (see,
for instance, \cite{Abrikosov,16}).

\textbf{Note 5.1}.  If to enter expression $c/z $ "inside" of
expression of dispersion function we will receive expression for
dispersion function $h(z)$ from our article \cite {Lesskis}
$$
h(z)=\dfrac{c}{z}\lambda(z)=\dfrac{c}{z}-\dfrac{1}{2}\int\limits_{-1}^{1}
\dfrac{\mu z-\eta_1^2}{\mu-z}d\mu=$$$$=\dfrac{c}{z}-z-(z^2-\eta_1^2)
\dfrac{1}{2}\ln\dfrac{z-1}{z+1}.
$$

\begin{center}\bf
 6. EXPANSIONS IN THE TERMS OF EIGEN FUNCTIONS
\end{center}

We will seek for the solution of the system of equations (2.5)
and (2.6) with boundary conditions (2.7)--(2.9) in the form
of linear combi\-na\-tion of discrete eigen solutions of the
characteristic system and integral taken over continuous spectrum of
the system. Let us prove that the following theorem is true.

\textbf{Theorem 6.1}. \emph{System of equations (2.5) and (2.6) with
boundary conditions (3.1), (3.6) and (2.7) has a unique solution,
which can be presented as an expansion by eigen functions of the
characteristic system:}
$$
h(x,\mu)=\dfrac{E_\infty}{z_0}\mu+$$$$+\dfrac{E_0}{z_0}\Big[
\ch\dfrac{xz_0}{\eta_0}(F(\eta_0,\mu)+F(-\eta_0,\mu))+
\sh\dfrac{xz_0}{\eta_0}(-F(\eta_0,\mu)+F(-\eta_0,\mu))\Big]+$$$$+
\intl_{-1}^{1}\Big\{\ch\dfrac{z_0x}{\eta}
\Big[ F(\eta,\mu)+F(-\eta,\mu)\Big]+\sh\dfrac{xz_0}{\eta}
\Big[-F(\eta,\mu)+F(-\eta,\mu)\Big]\Big\}\dfrac{E(\eta)}{z_0}\,d\eta,
\eqno{(6.1)}
$$
$$
e(x)=E_\infty+2E_0\ch\dfrac{z_0x}{\eta}+2\intl_{-1}^{1}
\ch\dfrac{z_0x}{\eta}E(\eta)\,d\eta.
\eqno{(6.2)}
$$

\emph{Here} $E_0$ \emph{and} $E_\infty$ \emph{is unknown
coefficients corresponding to the disc\-re\-te spectrum} ($E_0$ \emph{is
the amplitude of Debay}, $E_1$ \emph{is the amplitude of Drude),
}$E(\eta)$ \emph{is unknown function, which is called as coefficient
of discrete spect\-rum}.

When $(\Omega,\varepsilon)\in D^-$ in expansions (6.1) and (6.2) we
should take $E_0=0$. 

Further we will consider the following case $(\Omega,\varepsilon)\in
D^+$.

Our purpose is to find the coefficient of the continuous spectrum,
coeffi\-ci\-ents of the discrete spectrum and to built expressions for
electron distri\-bu\-tion function at the plasma surface and electric
field.

\textbf{Proof.} 
Let us consider expansion (6.1), we will replace in it
$\mu$ on $-\mu$. Then we will substitute $h (a, \mu) $ and $h (a,-\mu) $ in
The equation $ \dfrac {1} {2} [h (a, \mu)-h (a,-\mu)] =0$.
After variety of transformations let us have
$$
E_\infty\mu+
E_0\ch\dfrac{az_0}{\eta_0}\Big[F(\eta_0,\mu)+F(-\eta_0,\mu)\Big]+
$$
$$
+\int\limits_{-1}^{1}\ch\dfrac{az_0}{\eta}[F(\eta,\mu)+F(-\eta,\mu)]
E(\eta)d\eta=0,\quad -1<\mu<1,
\eqno{(6.3)}
$$
or
$$
E_\infty\mu+E_0\ch\dfrac{az_0}{\eta_0}\varphi(\eta_0,\mu)+
2\int\limits_{-1}^{1}\ch\dfrac{az_0}{\eta}F(\eta,\mu)E(\eta)d\eta=0.
\eqno{(6.4)}
$$

Let us pass from integral Fredholm equation  (6.4) to
singular integral equation with Cauchy kernel, having substituted in
(6.4) obvious representation $F(\eta,\mu)$
$$
E_\infty\mu+E_0\ch\dfrac{az_0}{\eta_0}\varphi(\eta_0,\mu)+2\int\limits_{-1}^{1}
\dfrac{\mu\eta-\eta_1^2}{\eta-\mu}
E(\eta)\ch\Big(\dfrac{z_0a}{\eta}\Big)d\eta-$$$$-4c\dfrac{\lambda(\mu)}{\mu}
E(\mu)\ch\dfrac{z_0a}{\mu}=0,\qquad 0<\mu<1.
\eqno{(6.5)}
$$

It is easy to check up, that function
$$
M(z)=\int\limits_{-1}^{1}\dfrac{z\eta-\eta_1^2}{\eta-z}E(\eta)
\ch\dfrac{z_0a}{\eta}d\eta
\eqno{(6.6)}
$$
is odd. Besides, all members of the equation (6.5) are
odd on $\mu$.

Extending the function $E(\eta)$ into the interval $(-1,0)$ evenly,
so that $E(\eta)=E(-\eta)$, and extending
the equation into the interval $(-1;1)$ unevenly, we transform
the equation (6.5) to the form
$$
E_\infty\mu+E_0\ch\dfrac{az_0}{\eta_0}\varphi(\eta_0,\mu)+
2\int\limits_{-1}^{1}
\ch\Big(\dfrac{z_0a}{\eta}\Big)\dfrac{\mu\eta-\eta_1^2}{\eta-\mu}
E(\eta)d\eta-$$$$-4c\dfrac{\lambda(\mu)}{\mu}
\ch\Big(\dfrac{z_0a}{\mu}\Big)E(\mu)=0,\quad -1<\mu<1.
\eqno{(6.7)}
$$\medskip

Let us reduce the equation (6.7) to boundary value problem
of Riemann --- Hilbert.
For this purpose we will take formulas of Sohotsky for the auxiliary
functions $M(z)$ and dispersion function $\lambda(z)$
$$
M^+(\mu)-M^-(\mu)=2\pi i\ch\Big(\dfrac{z_0a}{\mu}\Big)(\mu^2-\eta_1^2)
E(\mu), \quad -1<\mu<1,
\eqno{(6.8)}
$$
$$
\dfrac{M^+(\mu)+M^-(\mu)}{2}=M(\mu),\qquad -1<\mu<1,
$$
where
$$
M(\mu)=\int\limits_{-1}^{1}\dfrac{\mu\eta-\eta_1^2}{\eta-\mu}
\ch\Big(\dfrac{z_0a}{\eta}\Big)E(\eta)d\eta,
$$
and last integral is understood as singular in sense
integral principal value on Cauchy, and
$$
\lambda^+(\mu)-\lambda^-(\mu)=-\dfrac{i\pi}{c}\mu(\mu^2-\eta_1^2),\qquad
-1<\mu<1,
$$
$$
\dfrac{\lambda^+(\mu)+\lambda^-(\mu)}{2}=\lambda(\mu),\qquad
-1<\mu<1,
$$
where
$$
\lambda(\mu)=1-\dfrac{\mu}{2c}\int\limits_{-1}^{1}\dfrac{\mu'\mu-\eta_1^2}
{\mu'-\mu}d\mu'.
$$

As a result of use of last formulas we will come to the boundary value
problem
$$
E_\infty\mu+E_0\ch\dfrac{az_0}{\eta_0}\varphi(\eta_0,\mu)+
\big[M^+(\mu)+M^-(\mu)\big]+$$$$+
\dfrac{\lambda^+(\mu)+\lambda^-(\mu)}{\lambda^+(\mu)-
\lambda^-(\mu)}\big[M^+(\mu)-M^-(\mu)\big]=0.
$$

Let us transform this equation to the form
$$
(M^++M^-)(\lambda^+-\lambda^-)+(M^+-M^-)
(\lambda^++\lambda^-)+
$$
$$
+(\lambda^+-\lambda^-)(E_\infty\mu+
E_0\ch\dfrac{az_0}{\eta_0}\varphi(\eta_0,\mu)=0,\qquad -1<\mu<1.
$$

From here we receive the boundary condition of boundary value
problem of  Riemann --- Hilbert
$$
\lambda^+(\mu)
\Big[M^+(\mu)+\dfrac{E_\infty}{2}\mu+
\dfrac{E_0}{2}\ch\dfrac{az_0}{\eta_0}\varphi(\eta_0,\mu)\Big]-
$$
$$
=\lambda^-(\mu)\Big[M^-(\mu)+\dfrac{E_\infty}{2}\mu+
\dfrac{E_0}{2}\ch\dfrac{az_0}{\eta_0}\varphi(\eta_0,\mu)]=0, \quad
-1<\mu<1.
\eqno{(6.9)}
$$

Let us copy this problem in the form
$$
\Phi^+(\mu)-\Phi^-(\mu)=0, \quad -1<\mu<1.
\eqno{(6.10)}
$$
In the problem (6.10) $\Phi^{\pm}(\mu)$ is the boundary values in
interval $-1<\mu<1$ of function
$$
\Phi(z)= \lambda^+(z)\Big[M^+(z)+\dfrac{1}{2}E_\infty z+
\dfrac{1}{2}E_0\ch\dfrac{az_0}{\eta_0}\varphi(\eta_0,z)\Big],
$$
which is analytical function in complex plane with cut 
$\mathbb{C}\setminus [-1,1]$.

The problem (6.10) is the problem special case about jump.
It is problem of definition of analytical function on its jump 
on a contour $L$:
$$
\Phi^+(\mu)-\Phi^+(\mu)=\varphi(\mu), \qquad \mu\in L.
$$

The solution of such problems in a class decreasing at
infinitely remote point of functions is given by integral of Cauchy 
type
$$
\Phi(z)=\dfrac{1}{2\pi i}\int\limits_{L}\dfrac{\varphi(\tau)d\tau}
{\tau-z}.
$$

However, in the problem (6.10) unknown function $\Phi(z)$ has
at infinitely remote point following asymptotic
$$
\Phi(z)=O(z), \qquad z\to \infty.
$$

Therefore (6.11) it is necessary to search for the solution 
of the problem in the class of the growing
as $z $ at vicinity of infinitely remote point.

According to \cite{17} the general solution of the problem (6.10) 
is given by the formula
$
\Phi(z)=C_1z.
$

In explicit form the general solution of the problem (6.10) 
registers as follows:
$$
\lambda(z)\Big[M(z)+
\dfrac{1}{2}E_0\ch\dfrac{az_0}{\eta_0}\varphi(\eta_0,z)+
\dfrac{1}{2}E_\infty z\Big]=C_1z,
$$
where $C_1$ is the arbitrary constant.

From this general solution we can find function $M(z)$
$$
M(z)=-\dfrac{1}{2}E_\infty z-
\dfrac{1}{2}E_0\ch\dfrac{az_0}{\eta_0}\varphi(\eta_0,z)+
\dfrac{C_1z}{\lambda(z)}.
\eqno{(6.11)}
$$

Let us remove the pole at the solution (6.11) at infinitely remote point.
We receive, that
$$
C_1=\dfrac{1}{2}E_\infty\lambda_\infty.
\eqno{(6.12)}
$$

\begin{center}\bf
8. COEFFICIENTS OF THE CONTINUOUS AND DISCRETE SPECTRA
\end{center}

Now we will eliminate polar singularity at the solutiom (6.11) 
at points $\pm \eta_0$.
Let us allocate in the right part of the solution (6.11) 
members containing the polar singularity at points $z =\eta_0$. 
In the point vicinity $z =\eta_0$
taking into account equality $\lambda(\eta_0)=0$ fairly following 
expansion
$$
M(z)=-\dfrac{1}{2}E_\infty z-\dfrac{1}{2}E_0\ch\dfrac{az_0}{\eta_0}
\Bigg[\dfrac{\eta_0^2-\eta_1^2}{\eta_0-z}+\dfrac{\eta_0^2+\eta_1^2}
{\eta_0+z}\Bigg]+
$$
$$
+\dfrac{C_1\eta_0}{\lambda'(\eta_0)(z-\eta_0)+\cdots}.
$$

From here is visible, that for pole elimination at the point $z =\eta_0$
is necessary to equate to zero expression in a square bracket,
calculated at $z =\eta_0$. Then we receive, that:
$$
\ch\dfrac{az_0}{\eta_0}E_0=-\dfrac{E_\infty \lambda_\infty \eta_0}
{\lambda'(\eta_0)(\eta_1^2-\eta_0^2)}.
\eqno{(7.1)}
$$

The coefficient of continuous spectrum $E(\eta)$ we will find from 
the formula (6.9)
$$
E(\eta)\ch\dfrac{az_0}{\eta}=
\dfrac{1}{2\pi i}\cdot \dfrac{M^+(\eta)-M^-(\eta)}
{\eta^2-\eta_1^2}=
$$
$$
=\dfrac{ E_\infty\lambda_\infty}{2}
\dfrac{\eta}{2\pi i(\eta^2-\eta_1^2)}
\Big(\dfrac{1}{\lambda^+(\eta)}-\dfrac{1}{\lambda^-(\eta)}\Big).
\eqno{(7.2)}
$$

Let us take advantage of the boundary condition for electric field
$$
E_\infty+2E_0\ch\dfrac{az_0}{\eta_0}+2\int\limits_{-1}^{1}E(\eta)
\ch\dfrac{az_0}{\eta}d\eta=e_s.
\eqno{(7.3)}
$$

Let us substitute coefficient of the continuous spectrum (7.2) and 
coefficient of discrete spectrum (7.1) in the equation on electric field
(7.3). We will receive the following equation
$$
E_\infty-\dfrac{2E_\infty\lambda_\infty \eta_0}{\lambda'(\eta)(\eta_0^2-
\eta_1^2)}+E_\infty\lambda_\infty\dfrac{1}{2\pi
i}\int\limits_{-1}^{1} \Big(\dfrac{1}{\lambda^+(\eta)}-
\dfrac{1}{\lambda^-(\eta)}\Big)\dfrac{\eta d
\eta}{\eta^2-\eta_1^2}=e_s,
$$
or
$$
E_\infty\lambda_\infty\Bigg[\dfrac{1}{\lambda_\infty}-
\dfrac{2\eta_0}{\lambda'(\eta)(\eta_0^2-\eta_1^2)}+\dfrac{1}{2\pi
i}\int\limits_{-1}^{1} \Big(\dfrac{1}{\lambda^+(\eta)}-
\dfrac{1}{\lambda^-(\eta)}\Big)\dfrac{\eta d
\eta}{\eta^2-\eta_1^2}\Bigg]=e_s.
$$

Integral from this expression we will calculate by means of the 
contour integration and the theory of residual
$$
\int\limits_{-1}^{1} \Big(\dfrac{1}{\lambda^+(\eta)}-
\dfrac{1}{\lambda^-(\eta)}\Big)\dfrac{\eta
d\eta}{\eta^2-\eta_1^2}=$$$$=
\Big[\Res_{\infty}+\Res_{-\eta_1}+\Res_{\eta_1}+\Res_{\eta_0}
+\Res_{-\eta_0}\Big]\dfrac{z}{\lambda(z)(z^2-\eta_1^2)}=$$$$=
-\dfrac{1}{\lambda_\infty}+\dfrac{1}{\lambda_1}+
\dfrac{2\eta_0}{\lambda'(\eta)(\eta_0^2-\eta_1^2)}.
$$

Substituting this integral in the previous equation, we find that
$$
E_\infty=e_s\dfrac{\lambda_1}{\lambda_\infty}.
\eqno{(7.4)}
$$

Now by means of a relation (7.1) we find
$$
E_0\ch\dfrac{az_0}{\eta_0}=-e_s\dfrac{\lambda_1\eta_0}{\lambda'(\eta_0)
(\eta_0^2-\eta_1^2)}.
\eqno{(7.5)}
$$

By means of the found coefficients of discrete and continuous
spectra we find an electric field profile in the slab
$$
\dfrac{e(x)}{e_s}=\dfrac{\lambda_1}{\lambda_\infty}-
\dfrac{2\eta_0\lambda_1}{\lambda'(\eta_0)(\eta_0^2-\eta_1^2)}+$$$$+
\dfrac{\lambda_1}{2\pi i}\int\limits_{-1}^{1}\Big[\dfrac{1}
{\lambda^+(\eta)}-\dfrac{1}{\lambda^-(\eta)}\Big]
\dfrac{\ch(xz_0/\eta)}{\ch(az_0/\eta)}\dfrac{\eta
d\eta}{\eta^2-\eta_1^2}.
\eqno{(7.6)}
$$

Integral from (7.6) we will calculate by means of residual and contour
integration
$$
\dfrac{1}{2\pi i}\int\limits_{-1}^{1}\Big[\dfrac{1}
{\lambda^+(\eta)}-\dfrac{1}{\lambda^-(\eta)}\Big]
\dfrac{\ch(xz_0/\eta)}{\ch(az_0/\eta)}\dfrac{\eta
d\eta}{\eta^2-\eta_1^2}=
$$
$$
=\Big[\Res_{\infty}+\Res_{-\eta_1}+\Res_{\eta_1}+\Res_{\eta_0}
+\Res_{-\eta_0}+\sum\limits_{k=-\infty}^{+\infty}\Res_{t_k}\Big]
\dfrac{z\ch(xz_0/z)}{\sh(az_0/z)(z^2-\eta_1^2)\lambda(z)}=
$$
$$
=-\dfrac{1}{\lambda_\infty}+\dfrac{\ch(xz_0/\eta_1)}{\lambda_1
\ch(az_0/\eta_1)}+\dfrac{2\eta_0\ch(xz_0/\eta_0)}{\lambda'(\eta_0)
(\eta_0^2-\eta_1^2)\ch(az_0/\eta_0)}+$$$$+\dfrac{i}{az_0}
\sum\limits_{k=-\infty}^{+\infty}\dfrac{(-1)^kt_k^3\ch(xz_0/t_k)}
{\lambda(t_k)(t_k^2-\eta_1^2)}.
$$

Here $t_k$ are zeros of function $\ch(az_0/z)$,
$$
t_k=\dfrac{2az_0i}{\pi(2k+1)}, \qquad k=0,\pm 1, \pm 2, \cdots,
$$
also it has been considered that $\sh(az_0/t_k)=i(-1)^k$.

Hence, an electric field profile in the slab of degenerate plasma
is possible to express without quadratures
$$
\dfrac{e(x)}{e_s}=\dfrac{\ch(xz_0/\eta_1)}{\ch(az_0/\eta_1)}-
\dfrac{\lambda_1}{az_0}
\sum\limits_{k=-\infty}^{+\infty}\dfrac{t_k^3\ch(xz_0/t_k)}
{\lambda(t_k)(t_k^2-\eta_1^2)\sh(az_0/t_k)},
$$
or, that all the same,
$$
\dfrac{e(x)}{e_s}=\dfrac{\ch(xz_0/\eta_1)}{\ch(az_0/\eta_1)}+
\dfrac{i\lambda_1}{az_0}
\sum\limits_{k=-\infty}^{+\infty}\dfrac{(-1)^kt_k^3\ch(xz_0/t_k)}
{\lambda(t_k)(t_k^2-\eta_1^2)}.
\eqno{(7.7)}
$$

Here $\ch(xz_0/t_k)=\ch\dfrac{x}{a}(\pi k+\pi/2)$.

Integrating expression (7.7) on $x $ from $-a$ for $a$, we receive
$$
Q_1=\int\limits_{-a}^{a}e(x)dx=e_s
\dfrac{2\eta_1}{z_0}\th\dfrac{az_0}{\eta_1}-e_s
\dfrac{2\lambda_1}{az_0^2}\sum\limits_{k=-\infty}^{+\infty}
\dfrac{t_k^4}{\lambda(t_k)(t_k^2-\eta_1^2)}.
\eqno{(7.8)}
$$

Thus, all the coefficients of the expansions (6.1) and (6.2) are
found single-valued: the coefficient $E_\infty$ is determined
according to (7.4), the coefficient $E_0$ is determined according
to (7.5), the coefficient of the continuous spectrum $E(\eta)$ is
determined according to (7.2). Finding of the coefficients of
discrete and continuous spectra of the expansions (6.1) and (6.2)
completes the proving of existence of this expansions. Uniqueness of
the solution in the form of expansions (6.1) and (6.2) can be proved
easily with the use of contraposition method.

\begin{center}
  9. ENERGY ABSORPTION IN THE SLAB
\end{center}

The energy of an electromagnetic wave absorbed in the slab of
degenerate plasma is calculated under the formula
$$
Q=- \dfrac{\omega E_0}{8\pi}\int\limits_{-a}^{a}\Im E(x')\,dx'.
\eqno{(8.1)}
$$

Here dimensional quantities $E (x), x, a $ are written, and
$E(0)=E_0$.

Let us pass in (8.1) to dimensionless quantities
$$
e(x_1)=\dfrac{ev_F}{\nu \E_F}E(x), \quad x_1=\dfrac{x}{l}, \quad
e_s=\dfrac{ev_F}{\nu \E}E_0.
$$

It is as a result received
$$
Q=-\dfrac{\omega (\nu\E_F)^2le_s}{8\pi
(ev_F)^2}\Im \int\limits_{-a/l}^{a/l}e(x_1)dx_1.
\eqno{(8.2)}
$$

Further we will designate
$$
Q_1=\int\limits_{-a/l}^{a/l}e(x_1)dx_1.
$$

Then according to (8.2) we have
$$
Q=-\dfrac{\omega (\nu\E_F)^2le_s}{8\pi
(ev_F)^2}\Im Q_1.
\eqno{(8.2')}
$$

Besides, quantities $x_1$ and $a_1=a/l $ we will designate again through
$x$ and $a$.

Earlier for electric field which we will present in the form
$$
e(x)=E_\infty+2E_0\ch\dfrac{z_0x}{\eta_0}+\int\limits_{-1}^{1}
\ch\dfrac{z_0x}{\eta}E(\eta)d\eta,
\eqno{(8.3)}
$$
coefficients of continuous and discrete spectra have been calculated 
in the explicit form.

Integrating (8.3), we receive
$$
Q_1=2aE_\infty+4E_0\dfrac{\eta_0}{z_0}\sh\dfrac{az_0}{\eta_0}+
\dfrac{2}{z_0}\int\limits_{-1}^{1}\eta
E(\eta)\sh\dfrac{az_0}{\eta}\,d\eta.
\eqno{(8.4)}
$$

Substituting in (8.4) coefficients of discrete and continuous
spectra, we receive the following expression for electric field
$$
\dfrac{e(x)}{e_s}=\dfrac{\lambda_1}{\lambda_\infty}-
\dfrac{4\lambda_1\eta_0^2\th(az_0/\eta_0)}{z_0\lambda'(\eta_0)(\eta_0^2-
\eta_1^2)}+$$$$+\dfrac{2\lambda_1}{z_0}\dfrac{1}{2\pi
i}\int\limits_{-1}^{1}
\Big[\dfrac{1}{\lambda^+(\eta)}-\dfrac{1}{\lambda^+(\eta)}\Big]
\dfrac{\eta^2\th(az_0/\eta)}{\eta^2-\eta_1^2}.
\eqno{(8.5)}
$$

Integral from last expression we will calculate by means of the contour
integration and the theory of residual
$$
\dfrac{1}{2\pi
i}\int\limits_{-1}^{1}
\Big[\dfrac{1}{\lambda^+(\eta)}-\dfrac{1}{\lambda^+(\eta)}\Big]
\dfrac{\eta^2\th(az_0/\eta)}{\eta^2-\eta_1^2}=
$$
$$
=\Big[\Res_{\infty}+\Res_{-\eta_1}+\Res_{\eta_1}+
\Res_{-\eta_0}+\Res_{\eta_0}+\sum\limits_{k=-\infty}^{+\infty}
\Res_{t_k}\Big]
\dfrac{z^2\th(az_0/z)}{\lambda(z)(z^2-\eta_1^2)}=
$$
$$
=-\dfrac{az_0}{\lambda_\infty}+
\dfrac{\eta_1}{\lambda_1}\th\dfrac{az_0}{\eta_1}+
\dfrac{2\eta_0^2\th(az_0/\eta_0)}{\lambda'(\eta_0)(\eta_0^2-\eta_1^2)}-
\dfrac{1}{az_0}\sum\limits_{k=-\infty}^{+\infty}\dfrac{t_k^4}
{\lambda(t_k)(t_k^2-\eta_1^2)},
\eqno{(8.6)}
$$
where points $t_k$ are entered earlier.

Substituting (8.6) in (8.5), we receive
$$
\dfrac{Q_1}{e_s}=\dfrac{2\eta_1}{z_0}\th\dfrac{az_0}{\eta_1}-
\dfrac{2\lambda_1}{az_0^2}\sum\limits_{k=-\infty}^{+\infty}
\dfrac{t_k^4}{\lambda(t_k)(t_k^2-\eta_1^2)},
\eqno{(8.7)}
$$
that is equivalent to expression (7.8).

\begin{center}\bf
  10. CONCLUSION
\end{center}

In the present paper the linearized problem of plasma oscillations
in slab (particularly, thin films) in external longitudinal 
alternating electric field is solved
analytically. Specular  boundary conditions of
electron reflection from the plasma boundary are considered.
Coefficients of continuous and discrete spectra of the problem are
found, and electron distribution function on the plasma boundary and
electric field are expressed in explicit form.

Separation of variables
leads to characteristic system of the equations. Its solution in
space of the generalized functions allows to find the eigen
solutions of initial system of equations of Boltzmann --- Vlasov ---
Maxwell, correspond to  continuous spectrum.

Then the discrete spectrum of this problem consisting of zero of the 
dispersion equations is investigated. Such zeros are three. 
One zero is infinite remote point of complex plane. 
It is correspond to the eigen solution "Drude mode" independent of boundary 
conditions. Others two points of discrete spectrum are differing 
with signs two zero of dispersion function. 
These zero are correspond to the eigen solution named "Debay mode".

It is found out, that on  plane of parametres of a problem
($\Omega, \varepsilon $), where 
$ \Omega =\omega/\omega_p, \varepsilon =\nu/\omega_p $, there is such 
domain $D^+$ (its exterior is domain $D^-$), such, that if a point 
$(\Omega, \varepsilon)\in D^-$ Debay mode is absent.

Under eigen solution of initial system its general solution isobtained.
By means of boundary conditions  in an explicit form 
expressions for coefficients of discrete and continuous spectra are found.
Then in explicit form (without quadratures) absorption quantity
of energy of electric field in  slab of degenerate plasmas is found.

\end{document}